\documentclass[prb,reprint,twocolumn,citeautoscript,groupedaddress,noeprint,nofootinbib]{revtex4-2}
\usepackage{amsmath}
\usepackage{bm}
\usepackage{graphicx,color}
\usepackage{stix}
\usepackage{subfigure}
\usepackage{array}
\usepackage{braket}
\usepackage{nicefrac}
\usepackage[usenames,dvipsnames]{xcolor}

\newcommand{\Tr}{\operatorname{Tr}}
\newcommand{\im}{\operatorname{Im}}

\newcommand{\dos}{\operatorname{DOS}}
\newcommand{\spred}{^\mathrm{pred}}

\newcommand*{\gw}{{\textit{G}\textsubscript{0}\textit{W}\textsubscript{0}}}
\newcommand*{\GW}{\textit{GW}}

\definecolor{myblue}{rgb}{0,0,1}
\usepackage[breaklinks=true,colorlinks=true,linkcolor=myblue,urlcolor=myblue,citecolor=myblue]{hyperref}
\usepackage{comment}

\usepackage{enumitem,kantlipsum}

\newcommand*{\figref}[2][]{%
  Fig.~\hyperref[{fig:#2}]{%
    \ref*{fig:#2}%
    \ifx\\#1\\%
    \else%
      #1%
    \fi%
    }%
}

\newcommand*{\figsref}[2][]{%
  Figs.~\hyperref[{fig:#2}]{%
    \ref*{fig:#2}%
    \ifx\\#1\\%
    \else%
      #1%
    \fi%
  }%
}

\newcommand*{\Eqref}[1]{Eq.~\eqref{#1}}

\begin{document}

\title{Machine Learning Many-Body Green's Functions for Molecular Excitation Spectra}
\author{Christian Venturella}
\author{Christopher Hillenbrand}
\author{Jiachen Li}
\author{Tianyu Zhu}
\email{tianyu.zhu@yale.edu}
\affiliation{Department of Chemistry, Yale University, New Haven, CT, USA 06520}

\begin{abstract}
We present a machine learning (ML) framework for predicting Green's functions of molecular systems, from which photoemission spectra and quasiparticle energies at quantum many-body level can be obtained. Kernel ridge regression is adopted to predict self-energy matrix elements on compact imaginary frequency grids from static and dynamical mean-field electronic features, which gives direct access to real-frequency many-body Green's functions through analytic continuation and Dyson's equation. Feature and self-energy matrices are represented in a symmetry-adapted intrinsic atomic orbital plus projected atomic orbital basis to enforce rotational invariance. We demonstrate good transferability and high data efficiency of proposed ML method across molecular sizes and chemical species by showing accurate predictions of density of states (DOS) and quasiparticle energies at the level of many-body perturbation theory (\GW) or full configuration interaction. For the ML model trained on 48 out of 1995 molecules randomly sampled from the QM7 and QM9 datasets, we report the mean absolute errors of ML-predicted HOMO and LUMO energies to be 0.13 eV and 0.10 eV compared to \GW@PBE0. We further showcase the capability of this method by applying the same ML model to predict DOS for significantly larger organic molecules with up to 44 heavy atoms.
\end{abstract}

\maketitle

\section{Introduction}
Computational modeling of molecular excitation spectra plays a crucial role in revealing charge and energy transfer in light-matter interactions, understanding electron correlation in many-electron systems, and designing new optoelectronic materials and catalysts~\cite{Onida2002,Coropceanu2007,Nørskov2011,Zhu2018a}. Despite many developments in quantum chemistry and physics, accurate and fast first-principles prediction of photoemission spectra and quasiparticle energies of molecules and materials remains a significant challenge. While density functional theory (DFT)~\cite{Kohn1965} has been the workhorse for this task, molecular orbital energies from Kohn-Sham DFT are not true quasiparticle energies, and its accuracy heavily depends on underlying DFT functional~\cite{Perdew1982,Perdew1983,Perdew2017}. On the other hand, a rigorous and systematic solution to photoemission spectra beyond DFT can be obtained within the many-body Green's function (MBGF) framework~\cite{Hedin1965}. In recent years, promising \textit{ab initio} MBGF methods have been developed for molecules and periodic systems, based on many-body perturbation theory (\GW)~\cite{Hybertsen1986,Golze2019a,VanSetten2013,Ren2012,Govoni2015,Wilhelm2018,Zhu2021a,Lei2022,Li2022,Li2022a,Li2021c,Yeh2022a,Bintrim2021,Scott2023,Tolle2023}, second-order perturbation theory (GF2)~\cite{Phillips2014,Rusakov2016a,Hirata2015a,Hirata2017,Backhouse2020}, coupled-cluster theory (CC)~\cite{Nooijen1992,Nooijen1993,Zhu2019,Laughon2022,Bhaskaran-Nair2016,Peng2018a,Shee2019,Shee2022}, algebraic diagrammatic construction (ADC)~\cite{Banerjee2019,Banerjee2023}, density matrix renormalization group (DMRG)~\cite{Ronca2017a}, and quantum Monte Carlo~\cite{Gull2011}. Moreover, Green's function embedding methods including dynamical mean-field theory (DMFT)~\cite{Georges1996a,Kotliar2006,Zgid2011a,Zhu2020,Zhu2021c} and self-energy embedding theory (SEET)~\cite{Lan2015,Rusakov2019b,Iskakov2020} represent an efficient way to accelerate MBGF calculations of many-electron systems. In spite of these advances, the high computational scaling of MBGF methods prohibits their use in large-scale computational discovery of optoelectronic materials.

Data-driven machine learning (ML) has recently been employed to predict density of states (DOS) and quasiparticle energies at the DFT level from only atomic configurations~\cite{Ghosh2019,BenMahmoud2020,Fung2022,Singh2022,Yeo2019a}. Using thousands of (or more) discretized DOS on real frequency axis as training data, Gaussian process regression or deep neural network models are trained to predict DOS of organic molecules, bulk crystals, and amorphous materials, although the quality of results often depends on the resolution chosen to smooth the DOS~\cite{BenMahmoud2020}. A different ML approach is based on the SchNet model, where a latent Hamiltonian matrix is first predicted and molecular resonances are obtained as eigenvalues of the effective Hamiltonian~\cite{Schutt2019,Westermayr2021}. In the meantime, ML has also been explored to predict \GW~corrections to quasiparticle energy levels from DFT inputs~\cite{Westermayr2021,Caylak2021,Golze2022}. In addition, we note recent works in ML dielectric screening for accelerating \GW~and Bethe-Salpether equation (BSE) calculations~\cite{Dong2021a,Zauchner2023}. However, to the best of our knowledge, no current ML model predicts photoemission spectra at general quantum many-body level beyond independent-particle approximation and properly accounts for quasiparticle renormalization and satellites, which are important spectral features in correlated electron systems. Furthermore, to achieve high accuracy beyond DFT or even \GW, highly data-efficient ML models are needed, since MBGF calculations of full excitation spectra are usually very expensive.

In this work, we propose an ML framework for predicting many-body Green's functions (MLGF), where the training target is the self-energy matrix expressed on compact imaginary frequency axis grids calculated by any correlated quantum chemistry method, such as \GW~and full configuration interaction (FCI). We note that Green's function ML has been previously explored for solving single-site Anderson impurity models~\cite{Arsenault2014,Sturm2021,Sheridan2021}, but no such ML method exists for realistic molecules or materials. In our framework, kernel ridge regression (KRR) models are separately trained to predict diagonal and off-diagonal self-energy matrix elements on 18 imaginary frequency points, using static and dynamical mean-field electronic features such as the Fock matrix, density matrix, non-interacting Green's function, and hybridization function from Hartree-Fock (HF) or DFT calculations. A symmetry-adapted intrinsic atomic orbital plus projected atomic orbital (SAIAO) basis is introduced to ensure rotational invariance of feature and target matrices. We demonstrate the transferability and data efficiency of the MLGF method by predicting photoemission spectra and quasiparticle energies on water monomers and clusters, QM7 and QM9 datasets~\cite{Blum2009,Rupp2012a,Ruddigkeit2012,Ramakrishnan2015}, and polyacenes. Meanwhile, the MLGF approach provides important insights into orbital-resolved self-energy and correlation physics. Furthermore, we showcase its promise by predicting DOS of large organic semiconducting oligomers and drug molecules using ML model trained exclusively on small molecules.

\section{Method}
\subsection{Summary of Green's function theory}
The one-particle Green's function describes the propagation of an electron or a hole in many-particle systems ~\cite{Golze2019a}
\begin{equation}
     G(\mathbf{x}_1, t_1; \mathbf{x}_2, t_2) = -i \Braket{ N | \hat{T} [ \hat{\psi}(\mathbf{x}_1,t_1),  \hat{\psi}^\dagger(\mathbf{x}_2,t_2) ] | N },
\end{equation}
where $\hat{T}$ is time ordering operator, $\hat{\psi}$ and $\hat{\psi}^\dagger$ are annihilation and creation operators, $\ket{N}$ is the ground-state wave function, and $\{\mathbf{x}, t\}$ represents position and time. In the spectral re\-p\-re\-sen\-ta\-tion, the matrix elements of the Green's function are
\begin{subequations}
\begin{align}
 G^+_{pq}(z) &=  \Braket{ \Psi_0 | {a}_p  
    \left[z - (\hat{H}-E) \right]^{-1}  {a}^\dag_q | \Psi_0} , \\
  G^-_{pq}(z) &= \Braket{ \Psi_0 | {a}^\dag_q  
    \left[z + (\hat{H}-E) \right]^{-1}  {a}_p | \Psi_0}
\end{align}
\label{eq:exactgf}%
\end{subequations}
where $G^+(z)$ and $G^-(z)$ are addition (EA) and removal (IP) parts of Green's function, $|\Psi_0\rangle$ is the ground-state wave function, $\hat{H}$ is the Hamiltonian, $E$ is the ground-state energy, and $a_p$ and $a_q^\dag$ are annihilation and creation operators on orbitals $p$ and $q$. Here, $z$ is either real frequency with a small broadening $\eta$ ($z = \omega \pm i\eta$) or imaginary frequency at the Fermi level ($z = \epsilon_F + i\omega$). Throughout this paper, we will use $G(\omega)$ and $G(i\omega)$ to denote Green's functions on real and imaginary frequency axes, respectively. When computed on the real axis, the Green's function gives direct access to the spectral function
\begin{equation}
    A(\omega) \equiv -\frac{1}{\pi} \im \left[G(\omega)\right]
\end{equation}
and density of states (i.e., photoemission spectrum)
\begin{equation}
    \dos (\omega) \equiv \Tr A(\omega).
\end{equation}

The interacting (many-body) Green's function and non-interacting (mean-field) Green's function are related through Dyson's equation:
\begin{equation}
    G^{-1}(\omega) = G_0^{-1}(\omega) - \Sigma(\omega),
\end{equation}
where $\Sigma(\omega)$ is the self-energy, which is a frequency-dependent (i.e., dynamical) potential that captures many-body electron correlation. We propose to adopt self-energy as an ML target as it naturally provides an opportunity for efficient $\Delta$-ML from mean-field theories to correlated quantum chemistry methods. Moreover, Green's function and self-energy are more compact representation of many-body physics compared to wave function parametrizations as in CC or CI. 

\subsection{Feature and target representations}
Machine learning self-energy is challenging because it is a continuous matrix representation on the frequency (or time) axis. Previous ML works explored learning either a scalar-valued dynamical quantity (e.g., Green's function of a single orbital)~\cite{Arsenault2014} or a matrix-valued static quantity (e.g., mean-field Hamiltonian or coupled-cluster amplitude)~\cite{Schutt2019,Westermayr2021,Townsend2019}, but none of them targets a matrix-valued dynamical quantity. The success of the MLGF method thus relies on a compact representation of Green's function and self-energy. Because self-energy is much smoother on the imaginary frequency axis, we employ $\Sigma(i\omega)$ on a modified Gauss-Legendre grid as our training target~\cite{Ren2012,Zhu2021a}. $N_\omega=18$ grid points are used throughout this paper, meaning that $\Sigma(i\omega)$ is an $N_\mathrm{AO} \times N_\mathrm{AO} \times N_\omega$ matrix for an $N_\mathrm{AO}$-orbital molecule. From ML-predicted $\Sigma\spred(i\omega)$, real-axis self-energy $\Sigma\spred (\omega)$ is obtained through 18-point Pad\'e analytic continuation (AC)~\cite{Vidberg1977}. Finally, the many-body Green's function is calculated through Dyson's equation
\begin{equation}
    G\spred (\omega) = \left[G_0^{-1}(\omega) - \Sigma\spred (\omega)\right]^{-1}.
\end{equation}

We employ a symmetry-adapted intrinsic atomic orbital (IAO) plus projected atomic orbital (PAO) basis to represent the self-energy and mean-field feature matrices, which we will refer to as SAIAO basis. The IAO basis is first constructed by projecting occupied molecular orbitals from mean-field calculations onto a pre-defined minimal atomic orbital basis~\cite{Knizia2013d}, after which the PAO basis is constructed so that IAO+PAO basis spans the same Hilbert space as the original computational AO basis~\cite{Cui2020,Zhu2020}. The IAO+PAO basis is expected to enable more efficient MLGF model since it is already adapted for chemical environment and contains occupied mean-field information. To ensure rotational invariance, we apply a symmetry adaptation procedure similar to Ref.~\citenum{Qiao2020}. From the mean-field Fock matrix expressed in the IAO+PAO basis, we construct the SAIAO basis by diagonalizing small Fock matrix blocks $F^A_{nl}$ for each shell $(n,l)$ on every atom $A$:
\begin{equation}
    F^A_{nl} Y^A_{nl} = Y^A_{nl} \lambda^A_{nl},
    \label{eq:saiao}
\end{equation}
 where $n$ and $l$ are principal and angular quantum numbers, and $Y^A_{nl}$ and $\lambda^A_{nl}$ are the eigenvectors and eigenvalues of $F^A_{nl}$. The $Y^A_{nl}$ blocks, each of size $(2l+1)\times (2l+1)$, are combined into an orthogonal, block-diagonal transformation $Y$ that maps the IAO+PAO basis to the SAIAO basis
\begin{equation}
    \Ket{ \phi_i^\mathrm{SAIAO} } = \sum_p Y_{pi} \Ket{ \phi_p^\mathrm{IAO+PAO} } .
\end{equation}
Unlike Ref.~\citenum{Qiao2020}, we use the Fock matrix instead of the density matrix in \Eqref{eq:saiao}. Since the density matrix in the IAO+PAO basis, or any large AO basis, has near-zero matrix elements for high virtual orbitals, \Eqref{eq:saiao} would otherwise be ill-posed. 

We separately train ML models for diagonal and off-diagonal self-energy elements in SAIAO basis on each of 18 imaginary frequency points $\{i\omega_n\}$
\begin{align}
    \Sigma\spred_{ii}(i\omega_n) &= g_n^\mathrm{on}(\mathbf{f}^\mathrm{on}), \\
    \Sigma\spred_{ij}(i\omega_n) &= g_n^\mathrm{off}(\mathbf{f}^\mathrm{off}).
\end{align}
Here, $g_n^\mathrm{on}$ and $g_n^\mathrm{off}$ refer to ML models that predict diagonal and off-diagonal self-energy matrix elements at frequency $i\omega_n$, while $\mathbf{f}^\mathrm{on}$ and $\mathbf{f}^\mathrm{off}$ are corresponding feature vectors. This choice is motivated by the fact that diagonal self-energy elements typically have much larger magnitudes than off-diagonal self-energy elements, so it is more important to predict diagonal self-energy elements accurately. Separately training ML models for diagonal and off-diagonal self-energy elements allows more flexibility in the numbers of training data and features for capturing local and non-local correlation effects.

Inspired by MOB-ML~\cite{Welborn2018} and OrbNet~\cite{Qiao2022}, we employ electronic matrices from mean-field (HF or DFT) calculations as ML features, including core Hamiltonian matrix $h$, Fock matrix $F$, Coulomb matrix $J$, exchange-correlation potential matrix $K$, and density matrix $\gamma$. In contrast to these frequency-independent (i.e., static) features, we further introduce two dynamical features: the mean-field Green's function $G_0(i\omega)$ and the hybridization function $\Delta(i\omega)$ on imaginary frequency axis. The latter describes the delocalized environmental effect on a chosen subset of orbitals and is key to the success of Green's function embedding methods such as DMFT~\cite{Zhu2020,Zhu2021c}. In feature vector $\mathbf{f}^\mathrm{on}$, we define $\Delta_{ii}(i\omega)$ as
\begin{equation}
    \Delta_{ii} (i\omega) = \epsilon_F + i\omega - F_{ii} - [G_0(i\omega)]_{ii}^{-1}.
\end{equation}
In feature vector $\mathbf{f}^\mathrm{off}$, we take $2\times 2$ $ij$-block of $F$ and $G_0(i\omega)$ and then calculate $\Delta_{ij}(i\omega)$ as
\begin{equation}
    \Delta_{ij} (i\omega) = \left\{ \epsilon_F + i\omega - F_{ij} - [G_0(i\omega)]_\mathrm{sub}^{-1} \right\}_{01},
\end{equation}
where $[G_0(i\omega)]_\mathrm{sub}$ is
\begin{equation}
    (G_0)_\mathrm{sub} \equiv \begin{bmatrix}
    (G_0)_{ii} & (G_0)_{ij} \\
    (G_0)_{ji} & (G_0)_{jj} 
    \end{bmatrix}.
\end{equation}

In this work, we choose $G_0(i\omega_k)$ and $\Delta(i\omega_k)$ values on six frequency points: $\omega_k=[10^{-3}, 0.1, 0.2, 0.5, 1.0, 2.0]$ a.u., as ML features. This choice is motivated by observing the frequency dependence in Green's function and hybridization function of small molecular systems, but we do not attempt to optimize the number or values of these frequency points in this work. A preliminary test on the sensitivity of ML predictions to the choice of feature frequency points can be found in the SI. Since $G_0(i\omega_k)$ and $\Delta(i\omega_k)$ are complex matrices, we separately use their real and imaginary parts in feature vectors. In summary, the feature vectors in diagonal and off-diagonal ML models are
\begin{align}
    \mathbf{f}^\mathrm{on} &= \left[h_{ii}, F_{ii}, J_{ii}, K_{ii}, \gamma_{ii}, [G_0]_{ii}(i\omega_k), \Delta_{ii}(i\omega_k)\right], \label{eq:feature1} \\
    \mathbf{f}^\mathrm{off} &= \left[h_{ij}, F_{ij}, J_{ij}, K_{ij}, \gamma_{ij}, [G_0]_{ij}(i\omega_k), \Delta_{ij}(i\omega_k)\right].
    \label{eq:feature2}
\end{align}
All feature matrices are rotated to the SAIAO basis, and there are 29 features in each feature vector. We emphasize that this feature design together with the SAIAO basis guarantee equivariance in our ML method.

\subsection{ML workflow and computational details}
For both diagonal and off-diagonal regression models, we apply a min-max normalization to our features $\mathbf{f}^\mathrm{on}$ and $\mathbf{f}^\mathrm{off}$ and their respective targets $\Sigma_{ii}(i\omega_n)$ and $\Sigma_{ij}(i\omega_n)$. For the kernel in KRR, we use a linear combination of three Mat\'ern covariance functions $k_m$ plus a constant of 1:
\begin{equation}
K(\mathbf{f}_1, \mathbf{f}_2) = 1 + \sum^3_{m=1} k_m\left(\mathbf{f}_1, \mathbf{f}_2; l_m, \sigma_m^2, \nu = 3/2\right)
\end{equation}
Each $k_m$ is parametrized by length scale $l_m$, variance $\sigma^2_k$, and $\nu$ which controls differentiability and is fixed at 3/2:
\begin{align}
    k_m\left(\mathbf{f}_1, \mathbf{f}_2; l_m, \sigma_m^2 \right) = \sigma_m^2\Big (1+\frac{\sqrt{3}d_{12}}{l_m} \Big )\exp\Big({-\frac{\sqrt{3}d_{12}}{l_m}}\Big) .
\end{align}
where $d_{12}$ is the Euclidean distance between $\mathbf{f}_1$ and $\mathbf{f}_2$. The full kernel matrix $\mathbf{K}$ is constructed by computing $K(\mathbf{f}_i, \mathbf{f}_j)$ for all pairs of data points in our feature space. For this work, we choose $l_m = \{1.00, 0.50, 0.10\}$ and $\sigma^2_m = \{1.00, 0.50, 0.25\}$, foregoing any data-specific hyperparameter optimization. To train each prediction function for diagonal or off-diagonal elements, we fit independent real-valued kernels to each real and imaginary component of self-energy on 18 frequency points. With indices $i$ and $n$ referring to the $N_d$ data points and $N_y$ self-energy ML targets respectively, there are 36 copies of the kernel matrix $\mathbf{K}$ (diagonal and off-diagonal elements respectively), one for each self-energy target $y_n$, each copy corresponding to regression coefficients $\alpha_{in} \in \mathbb{R}^{N_d \times N_y}$ (i.e. the weight vectors). $N_d$ is the number of diagonal or off-diagonal data points and $N_y = 36$ is the effective dimension of our self-energy target. The final prediction functions $g_n(\mathbf{f})$ are obtained by minimizing the sum of squared residuals and an $L_2$ penalty term $\lambda \|\alpha\|^2$, giving the standard solution for KRR:
\begin{align}
    g_n(\mathbf{f}) &= \sum_{i=1}^{N_d} \alpha_{in} K(\mathbf{f}_i, \mathbf{f}), \\
    \alpha_{in} &= \left[(\mathbf{K}+\lambda \mathbf{I})^{-1}y_n\right]_i.
\end{align}
To achieve a close fit to the training data, we use a small regularization parameter of $\lambda = 10^{-5}$. Owing to $\lambda$ and that the Mat\'ern kernel is only $\lceil \nu \rceil - 1 = 1$ times differentiable, the matrix $\mathbf{K}+\lambda \mathbf{I}$ is well conditioned, with an inverse that can be computed efficiently with a Cholesky solver. The entire ML workflow is implemented using the scikit-learn package~\cite{scikitlearn}.

All Hartree-Fock, DFT, and FCI calculations were performed using the PySCF quantum chemistry package~\cite{Sun2020b}. Green's function calculations including \GW~and FCI-GF and Green's function data processing were done using the fcDMFT code~\cite{Zhu2020,Zhu2021c,Zhu2021a}. The IAO+PAO basis was generated using the libDMET code~\cite{Cui2020}. Unless otherwise noted, we employed the cc-pVDZ basis set~\cite{Dunning1989,Woon1993} (and cc-pVDZ-RI auxiliary basis~\cite{Weigend2002a} in \GW) for all molecular calculations, while GTH-DZVP basis set and GTH-HF pseudopotential~\cite{Hartwigsen1998,Vandevondele2005} were used for calculations with periodic boundary condition. For molecular structures newly optimized in this work, we used PySCF and the geomeTRIC package~\cite{geometric} to obtain ground-state geometries. 

Unless otherwise noted, ML predicted self-energy $\Sigma\spred (i\omega)$ was first rotated from SAIAO to molecular orbital (MO) basis and then analytically continued onto 201 real-frequency points from $-$1 a.u. to $+$1 a.u. with Pad\'e approximation and a broadening factor of $\eta = 0.01$ a.u. to obtain the real-axis self-energy $\Sigma\spred(\omega)$. In all \GW~calculations, the real-axis Green's function was obtained from $\Sigma\spred(\omega)$ by a diagonal approximation of Dyson's equation. i.e., only the diagonal self-energy in MO basis is used for calculating \GW~Green's function and quasiparticle energy, which is common for \GW~calculations. All \GW~calculations are one-shot \gw, where the quasiparticle energies are obtained from self-consistently solving the quasiparticle equation. For FCI-GF calculations, we used the full Dyson's equation without approximations. \\~\\

\onecolumngrid
\begin{figure*}[htb!]
    \centering
    \includegraphics[width=0.95\textwidth]{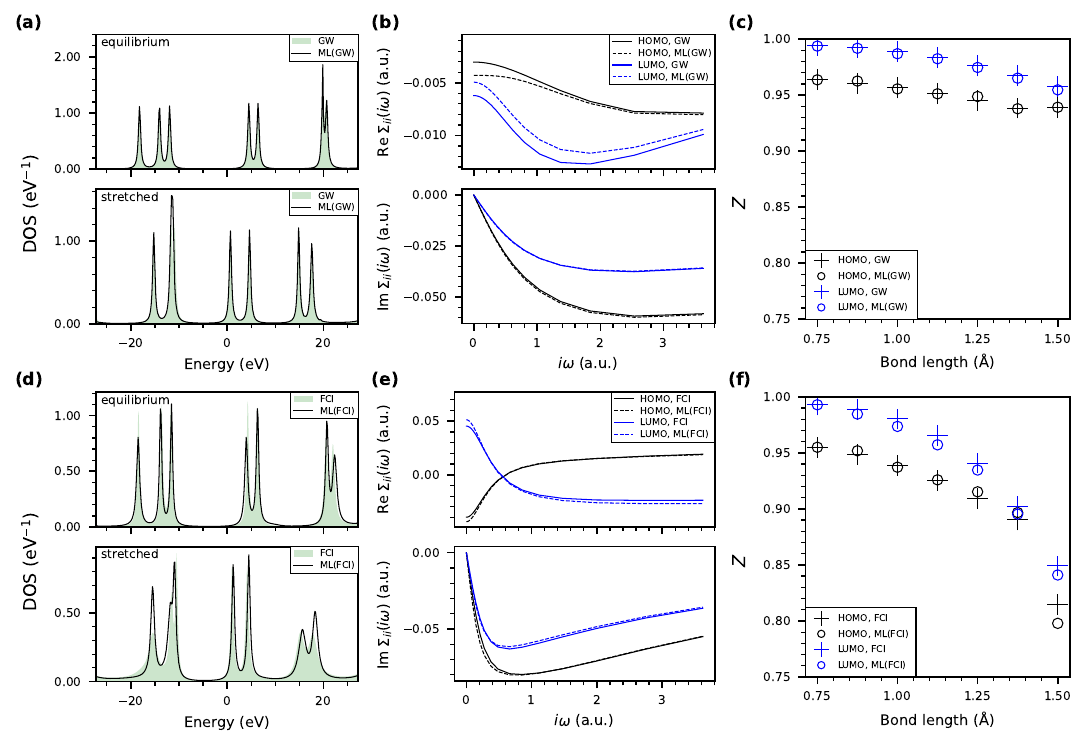}
    \caption{MLGF results for predicting water monomers at \GW@HF (a,b,c) and FCI levels (d, e, f). (a, d) MLGF-predicted DOS compared to the corresponding true DOS from \GW@HF and FCI for the equilibrium geometry and a stretched geometry. (b, e) For the stretched geometry, MLGF predictions of the complex-valued self-energy for HOMO and LUMO compared to the true self-energy on imaginary frequency axis. \mbox{(c, f)} MLGF-predicted quasiparticle renormalization $Z$ compared to true (\GW@HF or FCI) $Z$ plotted along symmetrically stretched bond length for near-equilibrium bond angle $\theta = 108^{\circ}$.} 
    \label{fig:1}
\end{figure*}
\twocolumngrid

\section{Results and Discussion}
\subsection{Water Monomers}
We first benchmarked MLGF on a set of 344 artificially generated water monomer geometries, expected to exhibit varying degrees of electron correlation. \GW@HF and FCI were adopted for generating training data, while HF was used for computing mean-field features and non-interacting Green's function. All except one of these geometries were generated by sampling $r_1+r_2$, $r_1-r_2$, and $\theta$  on a $7\times 7\times 7$ linearly-spaced grid where $r_1$ and $r_2$ are the two O--H distances in {\AA} and $\theta$ is the H--O--H bond angle; the last was an equilibrium geometry calculated at FCI level. For this work, we used $r_1+r_2 \in [1.5, 3.0]$, $r_1-r_2 \in [0.0, 0.5]$, and $\theta \in [72^{\circ}, 144^{\circ}]$.

All FCI calculations were done with a truncated cc-pVDZ basis (all $d$ orbitals removed). Because Pad\'e AC of the FCI self-energy was numerically unstable, we instead used adaptive Antoulas-Anderson (AAA) rational approximation~\cite{Nakatsukasa2018,Hofreither2021} on the imaginary-frequency FCI Green's function
\begin{equation}
    G\spred (i\omega) = \left[ G_0^{-1}(i\omega) - \Sigma\spred (i\omega) \right]^{-1}
\end{equation}
to obtain the FCI DOS. We note that the AAA algorithm may lead to unphysical Green’s functions and self-energies, although not found in this work. More robust analytic continuation methods will be tested in the future~\cite{Huang2023a}. To be consistent with FCI, we used full Dyson inversion for \GW@HF in this section.

Each MLGF model was trained on all available self-energy matrix elements from an evenly spaced $3\times 3\times 3$ subgrid $(r_1+r_2, r_1-r_2, \theta) \in \{1.5,2.25,3\} \times \{ 0, 0.25, 0.5\} \times \{72^\circ, 108^\circ, 144^\circ\}$. We calculated the coefficients of determination $R^2$ for self-energy predictions $\Sigma\spred (i\omega)$ in the SAIAO basis to first understand the ML performance on all 317 testing conformers. For \GW@HF predictions across all 36 diagonal/off-diagonal KRR models, the minimum and average $R^2$ values are 0.992/0.970 and 0.995/0.980 respectively, suggesting that off-diagonal self-energy matrix elements are more difficult to learn. For FCI, the ML performance is worse with minimum and average $R^2$ values for diagonal/off-diagonal self-energy predictions dropping to 0.901/0.854 and 0.963/0.923 respectively. We speculate that the worse ML performance for predicting the FCI self-energy is related to its larger magnitude and wider distribution on stretched water monomers compared to the \GW~self-energy, as demonstrated in Fig.~\ref{fig:1}b, c, e, f.

In \figref{1}, we present more detailed analyses on representative conformers in the testing set using \GW@HF or FCI training data. In \figref[a]{1} and \figref[d]{1}, we compare DOS predictions to each reference for configurations with weak electron correlation (equilibrium geometry from FCI optimization: $r_1 = r_2 = 0.99$, $\theta = 107.6^{\circ}$) and stronger electron correlation (stretched geometry: $r_1 = 1.58$, $r_2 = 1.17$, $\theta = 132.0^{\circ}$). ML prediction of DOS in \figref[a]{1} is in excellent agreement with the reference \GW@HF in terms of peak position and magnitude for both testing configurations. In \figref[d]{1}, MLGF accurately reproduces the main quasiparticle peaks around the Fermi level in the FCI DOS, while the prediction is slightly worse in the stretched geometry due to stronger correlation. Nevertheless, since MLGF learns the self-energy explicitly, it successfully captures the quasiparticle renormalization of peaks in the 10$\sim$20 eV region of the stretched geometry.

Focusing on frontier molecular orbitals, we find that our ML predictions of the HOMO (highest occupied molecular orbital) and LUMO (lowest unoccupied molecular orbital) self-energies on imaginary frequency axis are in good agreement with the true self-energies for the stretched geometry, as shown in \figref[b]{1} and \figref[e]{1}. We note that, the real part of \GW@HF self-energy has much smaller range compared to FCI self-energy, making an accurate KRR prediction harder, as the MO self-energy is obtained from mixing SAIAO self-energy matrix elements. However, this error negligibly affects the accuracy of DOS prediction (as seen in \figref[a]{1}), because small self-energy typically indicates small many-body correction to mean-field DOS.

We further display the range of electron correlation covered by our dataset by means of the quasiparticle renormalization factor of a given orbital $i$~\cite{Golze2019a}
\begin{equation}
    Z_i \equiv \Big[1 - \frac{\partial [ \im \Sigma_{ii}(i\omega)]}{\partial \omega} \Big|_{i\omega=0} \Big]^{-1},
\end{equation}
where possible values of $Z$ range from 0 to 1 and smaller $Z$ indicates stronger electron correlation. We calculate $Z$ for the HOMOs and LUMOs of symmetrically stretched monomers of fixed, near-equilibrium angle $\theta = 108^{\circ}$, which we plot against the bond length $r = r_1 = r_2$ in \figref[c]{1} and \figref[f]{1}. We find a good capability of MLGF to predict electron correlation from weak ($Z_\mathrm{FCI} > 0.98$) to stronger regimes ($Z_\mathrm{FCI} = 0.82$). Moreover, the signs of ML errors in $Z$ appear to be random, suggesting no clear bias for under-correlation or over-correlation. Seeing as $Z_\mathrm{GW}$ never drops below 0.94, these results exhibit the expected tendency of \GW~theory to under-correlate systems compared to FCI.  On the other hand, because \GW~is much more affordable for large molecules, we will use only \GW~training data for the remaining part of this paper, focusing on predicting DOS and quasiparticle energies of weakly correlated systems.

\subsection{Liquid Water}
After verifying the interpolation capability of MLGF in water monomer predictions, the size-transferability of MLGF was evaluated on four periodic water boxes consisting of 8, 16, 32, and 64 water molecules. Structures were obtained by equilibrating four independent \textit{ab initio} molecular dynamics (AIMD) trajectories in the NVT ensemble (300 K, 1.0 g/cm\textsuperscript{-3}) using the revPBE-D3 functional~\cite{Zhang1998a,Grimme2010a} and CP2K software package~\cite{Kuhne2020a}, following the procedure described in Ref.~\citenum{Zhu2021d}. One frame each was taken at 25 ps of equilibration. \GW@HF was employed to generate self-energy data for ML training and HF was used as the low-level mean-field theory.

To reduce the ML training dataset size for $\Sigma_{ij}(i\omega)$, matrix element samples are removed from training based on off-diagonal elements of the Coulomb matrix in the SAIAO basis. For water clusters, we remove samples when $|J_{ij}| < 5\times 10^{-3}$ a.u., as the off-diagonal self-energy element is typically negligible when two orbitals interact weakly. We trained three KRR models and validated their self-energy predictions on the largest 64-water snapshot. The first model was trained on a single 8-water cluster snapshot, referred to as the ``8'' model. The next model was augmented to also include a single 16-water cluster (``8+16''), and the final, largest model was further augmented with a 32-water cluster (``8+16+32'').

\begin{figure}[htb!]
    \centering
    \includegraphics[width=0.45\textwidth]{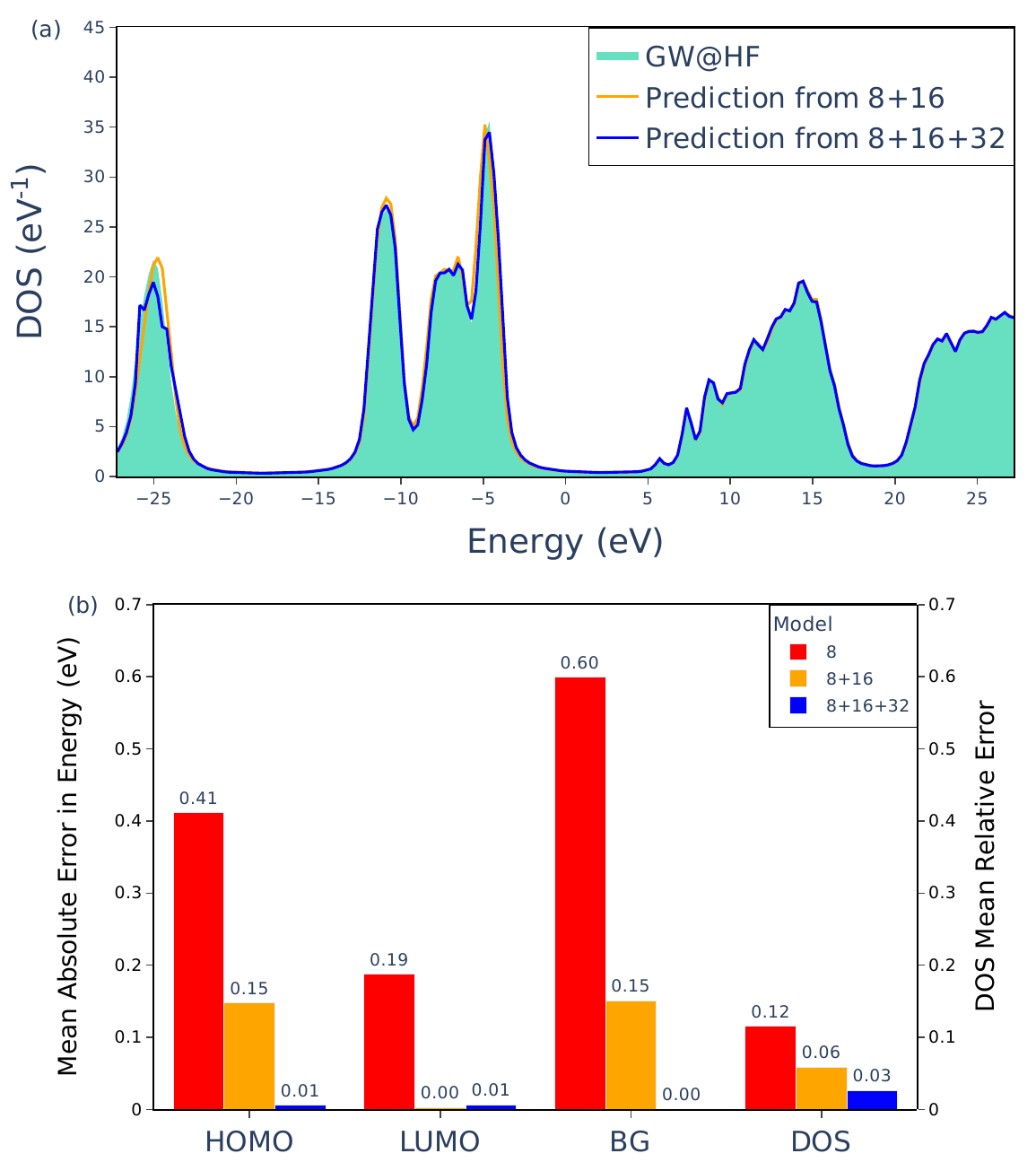}
    \caption{MLGF results for predicting periodic liquid water box of 64 water molecules. (a) ML predictions of DOS from a model trained on snapshots of 8 and 16 water molecules and a model trained on snapshots of 8, 16, and 32 water molecules. DOS is not shifted with respect to vacuum. (b) Error reductions in predicting HOMO energy, LUMO energy, band gap (BG), and DOS for 64-water cluster upon subsequent addition of 16-water and 32-water snapshots to the training data.} 
    \label{fig:2}
\end{figure}

As shown in \figref[a]{2}, the predicted 64-water DOS quickly converges to the \GW@HF reference DOS upon subsequent addition of 16 and 32 water clusters to the training data. The DOS prediction of the 64-water cluster from the ``8+16+32'' model is in near-perfect agreement with the true DOS for a wide range of frequency, while most of the discrepancy for both ``8+16'' and ``8+16+32'' models comes from the lower-lying occupied orbitals. To quantify the DOS error, we define a metric that measures the absolute relative DOS error on a real-frequency axis of $N_\omega=201$ points
\begin{equation}
    \epsilon(\dos_p) \equiv \frac{\sum_i^{N_\omega} \left|\dos_p(\omega_i) - \dos_t(\omega_i)\right|}{\sum_i^{N_\omega} \dos_t(\omega_i)} ,
\end{equation}
where $\dos_p$ and $\dos_t$ are the predicted and true DOS respectively. In \figref[b]{2}, we show that the relative DOS error drops from 0.12 in the ``8'' model to only 0.03 in ``8+16+32'' model.

In \figref[b]{2}, ML-predicted LUMO energy converges to the target value within $0.01$ eV after inclusion of 16-water clusters (``8+16'') without a need to include 32-water clusters in training. In contrast to the LUMO energy, the HOMO energy converges more slowly, approaching the true value within $0.01$ eV only after inclusion of the 32-water cluster in the training data (``8+16+32''). Similarly, the minimum error for the band gap (< 0.01 eV) is reached after inclusion of the 32-water cluster in training data. In summary, this benchmark suggests that MLGF is capable of predicting highly accurate photoemission spectrum of 64-water cluster using only one 32-water snapshot in ML training, which demonstrates its transferability across molecular cluster sizes. More detailed and unrounded errors for all models across predictions of all cluster sizes can be found in the Supporting Information (SI).

\twocolumngrid
\begin{figure*}[htb!]
\includegraphics[width=0.95\textwidth]{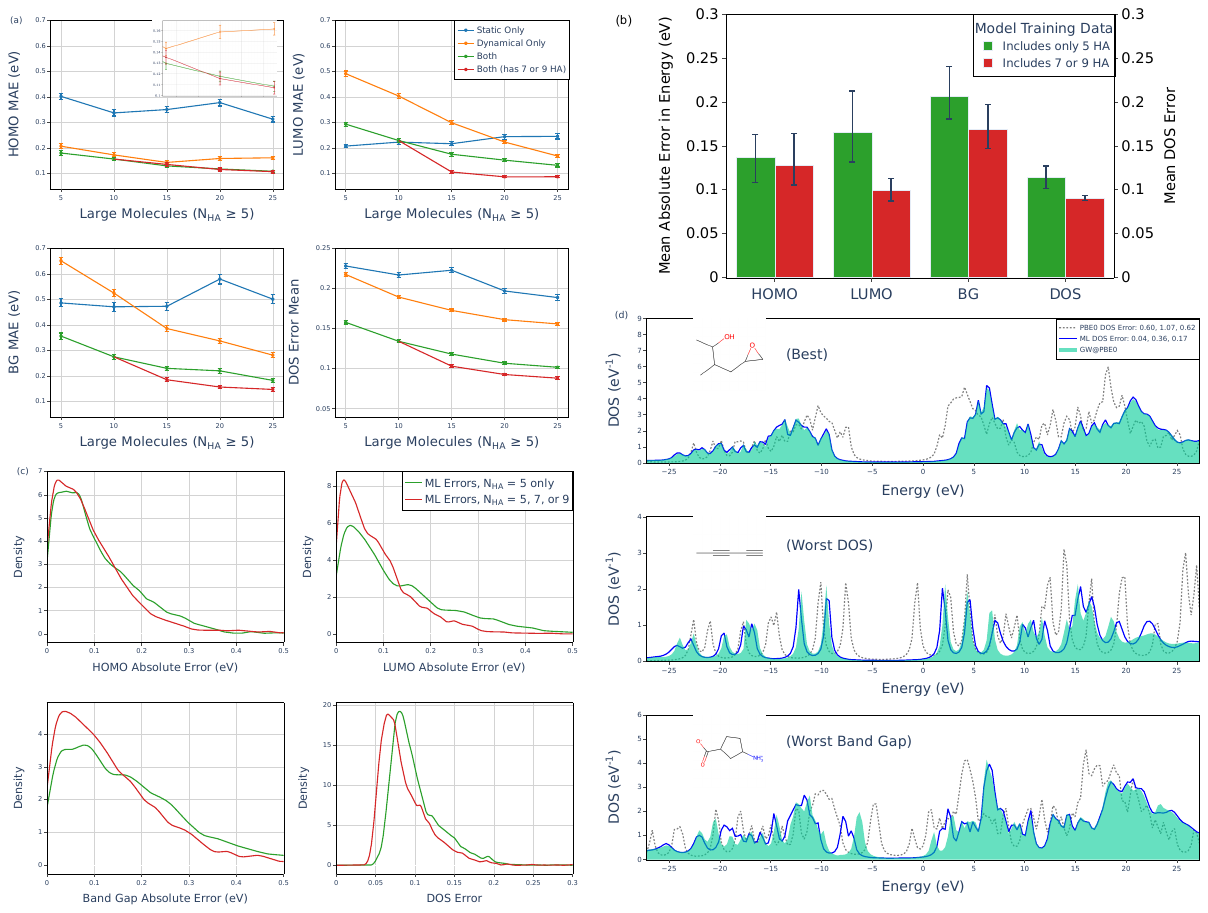}
\caption{MLGF results for QM7 and QM9 datasets. (a) Errors of HOMO energy, LUMO energy, band gap (BG), and DOS upon subsequent addition of 5 large molecules to the training data for best of 10 replicas. Error bars reflect a 95\% confidence internal for the mean errors within the best replica, with inset showing overlap in HOMO energy confidence intervals for largest models. ``Both'' refers to using both static and dynamical features as proposed in Eqs.~\ref{eq:feature1} and \ref{eq:feature2}. (b) Average of mean absolute errors across replicas for largest models (including 25 large molecules) for HOMO energy, LUMO energy, band gap (BG), and DOS. Top bars represent worst average error of all replicas, bottom bars represent best of all replicas. (c) Absolute error distributions (kernel density estimate) for HOMO energy, LUMO energy, band gap (BG), and DOS in the best model. (d) Example DOS predictions for smallest DOS error (``Best''), largest DOS error (``Worst DOS''), and largest band gap error (``Worst Band Gap'') from the best ML model. PBE0 DOS is also shown for reference.} 
\label{fig:3}
\end{figure*}

\subsection{QM7 and QM9 Molecules}
We now explore the capability of MLGF to generalize to different chemical species and larger molecular sizes simultaneously. We calculated \GW@PBE0 for 1,995 molecules from QM7 and QM9 datasets, including all molecules of five or fewer heavy atoms ($N_\mathrm{HA}\leq5$) from both datasets (395 molecules), and 800 of the largest molecules from each of QM7  ($N_\mathrm{HA}=7$) and QM9 ($N_\mathrm{HA}=9$). Since the \GW@PBE0 self-energy is our ML target, we used the PBE0 exchange-correlation potential for the static feature $K$.  As in periodic water, Coulomb screening is employed to reduce the sample sizes for learning $\Sigma_{ij}(i\omega)$, but with a looser threshold for excluding elements $|J_{ij}| < 5\times 10^{-2}$ a.u.


To understand the data efficiency and feature effectiveness of MLGF, we devised the following cross-validation experiments:
\begin{enumerate}[wide,itemsep=0pt]
    \item Start from a baseline model only trained with molecules in QM7 and QM9 having $N_\mathrm{HA}\leq3$ (a starting point of 23 molecules), whose computational cost for \GW@PBE0 is negligibly low.
    \item Successively add 5 random molecules having $N_\mathrm{HA}=5$ to the training data, up to a maximum 25 molecules having  $N_\mathrm{HA}=5$.
    \item From models with 10 molecules having  $N_\mathrm{HA}=5$, instead of adding more training molecules with $N_\mathrm{HA}=5$, add molecules having either $N_\mathrm{HA}=7$ or  $N_\mathrm{HA}=9$ into the training set.
    \item On training sets with only molecules having $N_\mathrm{HA}=5$, train models with static features only and models with dynamical features only.
\end{enumerate}
We performed Steps 1-4 for ten independent replicas, as shown in \figref{3}. For validations within each replica, the union of all training data splits in each experiment are considered seen by the model, which always reserves 1,932 molecules for testing (i.e., a 2:98 train-test split). A visual representation of the training examples for the MLGF model depicted in \figsref{3} and \ref{fig:4} can be found in the SI.

In \figref[a]{3}, the performance of each model in the best replica of our cross-validation experiments is plotted against increasing training data sizes. The error bars represent a 95\% confidence interval for the mean absolute error (MAE) of each model (within $\pm 2\sigma_\mu$ where $\sigma_\mu$ is the standard deviation of the sample mean). We find that using both static and dynamical features is almost always better than having static or dynamical features alone, which indicates that both static and dynamical features carry important and unique information for the self-energy prediction. Across all metrics, but most starkly for LUMO energy and band gap, training on only dynamical features converges more consistently upon training data augmentation compared to training on only static features. For LUMO prediction, static features provide a good baseline for 5 molecules with $N_\mathrm{HA}=5$ in the training data, but augmentation to 10, 15, 20, and 25 such molecules fails to reduce the error. This suggests that the LUMO energy is affected by differences between chemical environments which are better captured by dynamical features. Compared to the LUMO and band gap predictions, the HOMO energy prediction is harder to improve upon increasing training molecule sizes. Augmentation with molecules having $N_\mathrm{HA}>5$ has no clear benefit over molecules having $N_\mathrm{HA}=5$. 

In contrast to HOMO, LUMO, and band gap errors, we observe trends in the DOS error (\figref[a]{3}) that can be more clearly interpreted. Based on DOS error, we confirm that combining static and dynamical feature sets is better than using either feature set on its own. Meanwhile, as expected, augmenting the training data with molecules having $N_\mathrm{HA} > 5$ is better than restricting the training data to $N_\mathrm{HA} = 5$. Of the four metrics considered, the DOS error more stably and monotonically decreases because it represents an average error over all orbitals in the test set. 

Across all replicas' largest models in \figref[b]{3}, the average MAEs of the predicted HOMO energy, LUMO energy, and band gap are 0.13/0.10/0.17 eV respectively, while the best model gives MAEs of 0.11/0.09/0.15 eV. The LUMO prediction improves the most upon inclusion of molecules having $N_\mathrm{HA}=7$ or $N_\mathrm{HA}=9$ in the training data. However, there is considerable variance in model performance across MLGF replicas, especially in the prediction of the HOMO energy, where the errors for models with molecules having $N_\mathrm{HA}=5$ overlap with those having $N_\mathrm{HA}=7$ or $N_\mathrm{HA}=9$. Thus, the HOMO error convergence in \figref[a]{3} roughly reflects the typical behavior among all replicas. As in \figref[a]{3} for one replica, the DOS error is consistently improved upon inclusion of the largest molecules in QM7 and QM9. 

In \figref[c]{3}, absolute error distributions for the best replica are generated via Gaussian kernel density estimation with a bandwidth parameter of $0.10 \times \sigma$, where $\sigma$ is the sample standard deviation for each distribution. By all metrics, the model trained on molecules with $N_\mathrm{HA}=7$ or $N_\mathrm{HA}=9$ (red curve) perform better compared to the model restricted to molecules with $N_\mathrm{HA}=5$ (green curve); the absolute error mean is reduced and the distribution spread is also reduced. In \figsref[a--c]{3}, the HOMO error is the hardest to reduce upon training data augmentation, which suggests that adding more diverse chemical species and bonding types may be more important than adding larger molecules and more features may be necessary to distinguish frontier SAIAOs in larger datasets.

In \figref[d]{3}, we present specific molecular examples for the best and worst DOS predictions as well as the worst band gap prediction drawn from distributions in \figref[c]{3}. Clearly, even in the worst cases, the MLGF-predicted DOS is significantly better than the PBE0 DOS when using \GW@PBE0 as the reference. The best DOS prediction is nearly perfect, since an epxoide species similar to this test molecule exists in the training set (Fig.~S3). Given that the DOS error distribution in \figref[c]{3} is tightly centered around 0.065, marginally worse than the best test case DOS (0.04), most of the other DOS in this error distribution are of similarly good quality. In the worst DOS case, MLGF predicts the DOS near the band gap quite well, with most DOS error coming from high virtual orbitals. We also note that our DOS error metric inherently penalizes smaller molecules with sparse sharp peaks. For the molecule with the worst band gap prediction, ML predicts spectrum of virtual and core orbitals with good accuracy, despite compounding HOMO and LUMO errors that overestimate the band gap by several eV. We expect this for the unusual zwitterion in question, as our training examples lack an analogous charge-separated species. 

Finally, we employ principal components analysis (PCA) to understand the distribution of diagonal-element features, which can be found in Fig.~S2. From PCA, we discern that specific orbitals are distinctly clustered by SAIAO angular momentum ($s, p, d, f$) and by core, IAO, or PAO character. In addition, PCA roughly quantifies relative feature importance, highlighting varying contributions from our 29 diagonal SAIAO features to the first two principal components. Put into context with our previous results, PCA demonstrates that our SAIAO features can successfully express mean-field information across orbitals of varying character within a diversity of chemical structures. 

In summary, these results demonstrate that MLGF can successfully predict photoemission spectra at the \GW~level with good accuracy and capability to extrapolate in terms of molecule size and element composition, while maintaining high data efficiency. Our application of MLGF to QM7 and QM9 molecules has also exposed some challenges. Specifically, the HOMO errors are less straightforward to converge among the metrics we considered, and in general, MLGF accuracy is sensitive to the molecule samples seen during ML model training. Possible solutions to these aforementioned challenges are proposed in the Conclusion section. 

\subsection{Large Molecules and Polycyclic Aromatic Hydrocarbons}

\begin{figure}[hbt]
\centering
\includegraphics[width=0.45\textwidth]{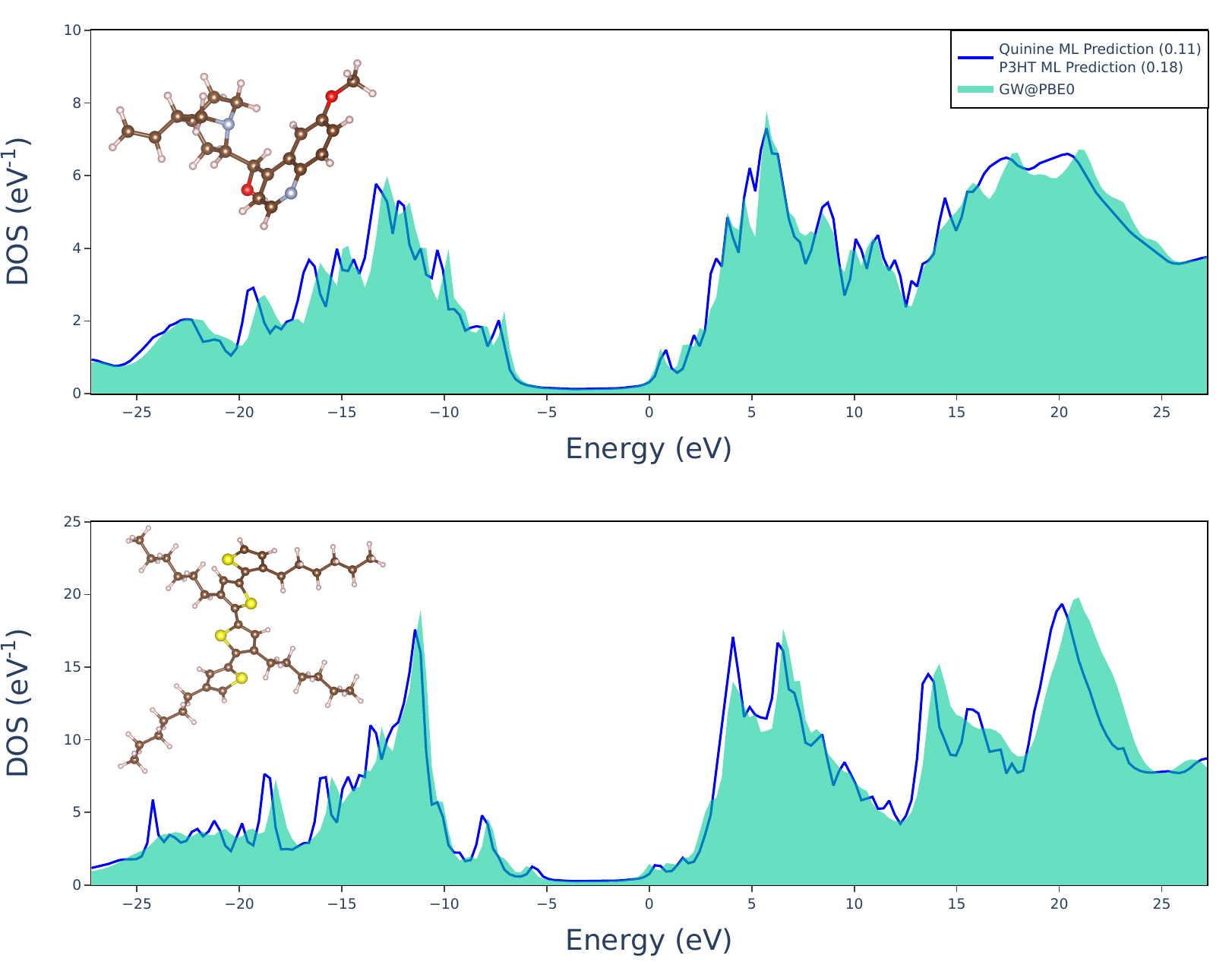}
\caption{MLGF prediction of quinine and P3HT trained on 48 molecules from QM7 and QM9. HOMO, LUMO, band gap errors for quinine (in eV): 0.18, 0.16, 0.34. HOMO, LUMO, band gap errors for P3HT (in eV): 0.28, 0.38, 0.10. } 
\label{fig:4}
\end{figure}

From our QM7 and QM9 replicas, we select a model trained on 25 molecules having $N_\mathrm{HA}\leq9$ to predict the DOS of two large molecules of photochemical relevance: quinine with $N_\mathrm{HA}=24$ and poly(3-hexylthiophene-2,5-diyl) (P3HT) oligomer (four 3HT units) with $N_\mathrm{HA}=44$. Out of ten available replicas, we select a model based on chemical intuition, which includes two aromatic sulfur heterocycles to capture the thiophene rings in P3HT, in addition to a bicylic nitrogen ring to roughly match the bicylic component of quinine (see Fig.~S4). With this model, we attain absolute errors of 0.18/0.16/0.34 eV for quinine and 0.28/0.38/0.10 eV for P3HT oligomer on HOMO energy, LUMO energy, and band gap, as shown in \figref{4}. In the meantime, both predicted DOS are in good agreement with the corresponding \GW@PBE0 DOS. In regard to our chemically inspired selection of an ideal model for this prediction task, we admit that MLGF performance can vary depending on which functional groups are seen in training. A more systematic selection of the training examples may be devised for a specific prediction task in the future.

\begin{figure*}[hbt!]
\centering
\includegraphics[width=0.8\textwidth]{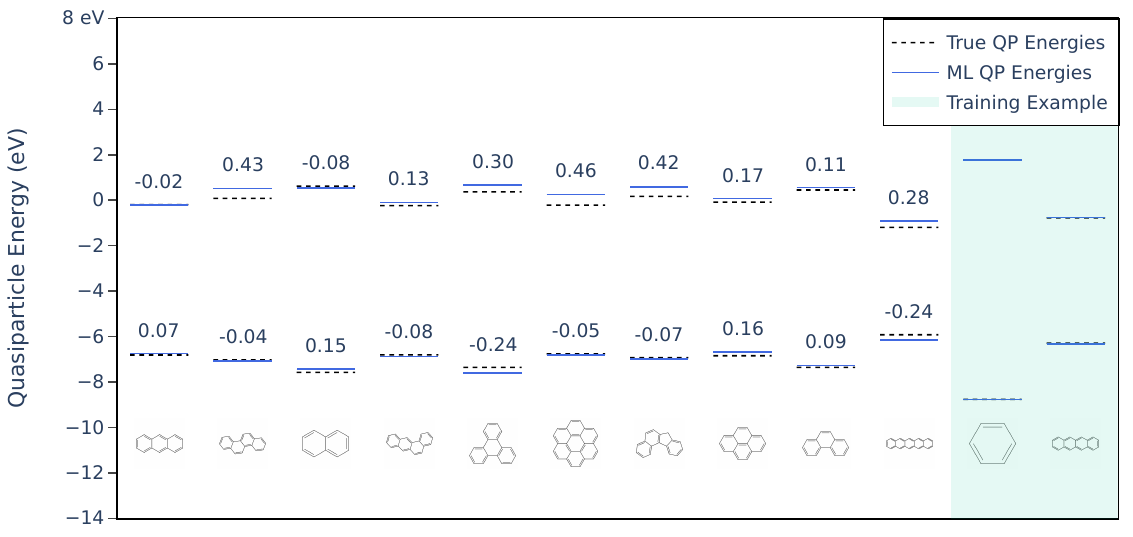}
\caption{MLGF prediction of HOMO and LUMO energies for PAHs with signed errors given above the corresponding energy levels.} 
\label{fig:5}
\end{figure*}

To test the capability of MLGF to predict highly delocalized electron correlation, quasiparticle energies were calculated for 11 polycylic aromatic hydrocarbons (PAHs) at \GW@PBE0 level with aug-cc-pVDZ basis set~\cite{Kendall1992a}. Benzene and tetracene were used as training examples and $\Sigma_{ij}(i\omega)$ elements were removed from training when $|J_{ij}| < 5\times 10^{-3}$ a.u. The predicted and true values for HOMO and LUMO energies are summarized in \figref{5}. 

Molecules with sizes between that of benzene and tetracene are interpolated with reasonable accuracy of $\pm 0.15$ eV for both HOMO and LUMO energies. Molecules larger than tetracene are predicted with larger errors, with the unsigned error of the LUMO errors usually larger than that of the HOMO. All band gap errors for molecules of size up to 24 carbon atoms (coronene) are below 0.6 eV. The difficulty of predicting conjugated systems demonstrates the limitation of MLGF. While current feature design in the SAIAO basis offers a local representation for transferability in terms of size and chemical composition, our feature set is less successful in extrapolating many-body physics when electrons are delocalized over larger distances than seen in training examples. \\~\\~\\

\onecolumngrid
\twocolumngrid

\section{Conclusions}
We have presented an ML framework for predicting photoemission spectra at quantum many-body level by learning the self-energy and Green's function from mean-field electronic features. This is achieved through a combination of compact self-energy representation on imaginary frequency axis, frequency-dependent dynamical features, rotation-invariant SAIAO local orbital basis, and orbital-specific KRR models. Our MLGF method accurately predicts the self-energy, DOS, and quasiparticle renormalization at the \GW~and FCI levels for both weakly and strongly correlated regimes of water monomers. We also demonstrate good transferability and high data efficiency across molecular sizes and chemical compositions in predicting DOS and quasiparticle energies for liquid water, diverse organic molecules, and polyacenes. We further show that the MLGF method is capable of predicting accurate DOS for certain large and complex photoactive molecules well beyond the molecular sizes utilized in ML training. From these test cases, we identify challenges in reducing quasiparticle energy errors below 0.1 eV when extrapolating system sizes, in particular near the band gap for large organic molecules and extended conjugated systems. 

Given that our initial studies mainly focus on creating compact electronic features, ML prediction post-processing, and benchmarking various chemical systems, we have yet to explore natural extensions of this work that could address the challenges we have seen thus far. For featurizing larger molecules from more chemically varied datasets and better incorporating non-local electron interactions, electronic features beyond single-orbital and two-orbital matrix elements may be explored. In addition, one may utilize categorical features that describe orbital character, such as those exposed in the PCA clustering of SAIAOs in QM7 and QM9. To augment our original set of 29 electronic features, a neural network architecture is likely needed to afford greater feature flexibility. Appreciating the sensitivity of a given prediction task to the specific training examples, active learning may be employed in cases where manual selection of chemical structures is non-intuitive or one needs a more automatic procedure for large-scale applications. These directions are currently being pursued in our group. 

\begin{acknowledgments}
This work was supported by a start-up fund from Yale University. J. Li acknowledges support from the Tony Massini Postdoctoral Fellowship in Data Science from Yale University. We thank the Yale Center for Research Computing for supercomputing resources. 
\end{acknowledgments}

\bibliography{mlgf}

\raggedbottom

\widetext
\clearpage
\begin{center}
\textbf{\large Supporting Information for: \\ Machine Learning Many-Body Green's Functions for Molecular Excitation Spectra}
\end{center}
\setcounter{equation}{0}
\setcounter{figure}{0}
\setcounter{table}{0}
\setcounter{page}{1}
\setcounter{section}{0}
\makeatletter
\renewcommand{\theequation}{S\arabic{equation}}
\renewcommand{\thefigure}{S\arabic{figure}}
\renewcommand{\thetable}{S\arabic{table}}
\renewcommand{\thesection}{S\arabic{section}}
\renewcommand{\bibnumfmt}[1]{[S#1]}
\renewcommand{\citenumfont}[1]{S#1} 

\section{Validation of Periodic Water Predictions}
We trained ML models on cluster sizes up to 32 waters, and validated them on all cluster sizes (up to 64 waters). Ocassionally we see slight worsening of prediction quality for smaller clusters upon augmenting the training data with larger clusters (for example the worsening DOS error for an 8-water cluster upon adding 16 and 32 water cluster). Considering we more frequently see smaller cluster prediction improve (for the HOMO/LUMO and band gap errors), we can regard these observations as not problematic.
\begin{figure}[hbt]
\centering
\includegraphics[width=0.52\textwidth]{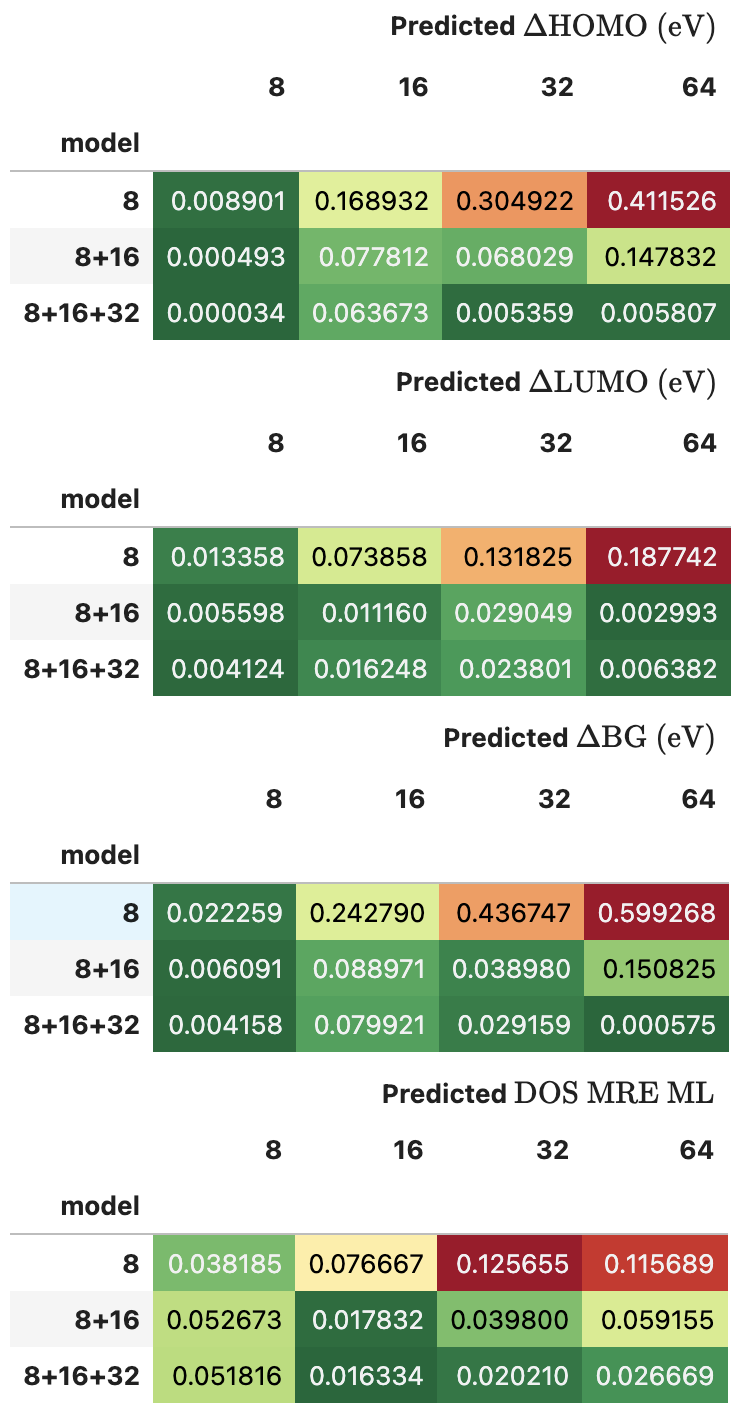}
\caption{Absolute errors for periodic water for all ML models (8, 8+16, 8+16+32) on all size clusters (8, 16, 32, 64).}
\label{fig:si_pw}
\end{figure}

\section{QM7 and QM9 Supplemental Analysis}
\begin{figure}[hbt]
\centering
\includegraphics[width=0.9\textwidth]{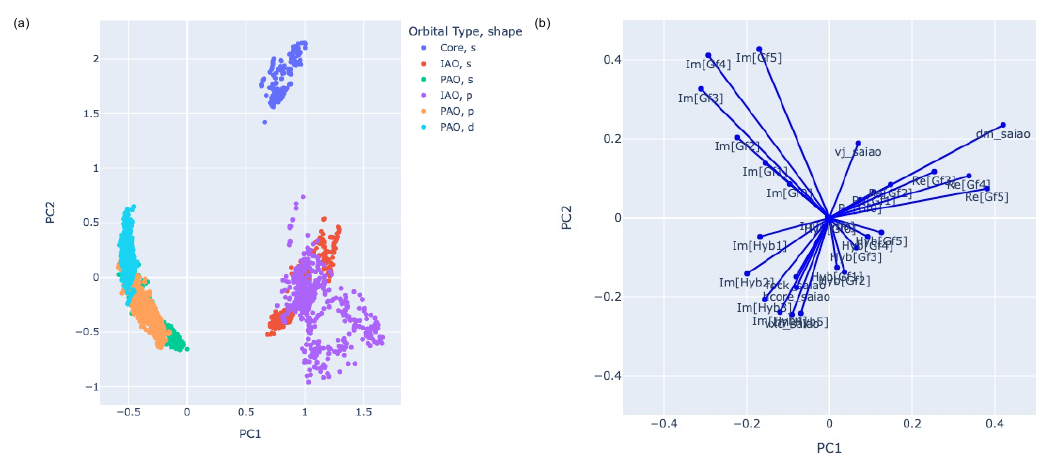}
\caption{PCA for best QM7 and QM9 replica from \figref{3}. (a) Principal components color coded by SAIAO type (by $s, p, d, f$) and core, IAO, PAO. (b) Biplot showing feature contributions to PC1 and PC2.}
\label{fig:si_pca1}
\end{figure}

\begin{figure}[hbt]
\centering
\includegraphics[width=0.95\textwidth]{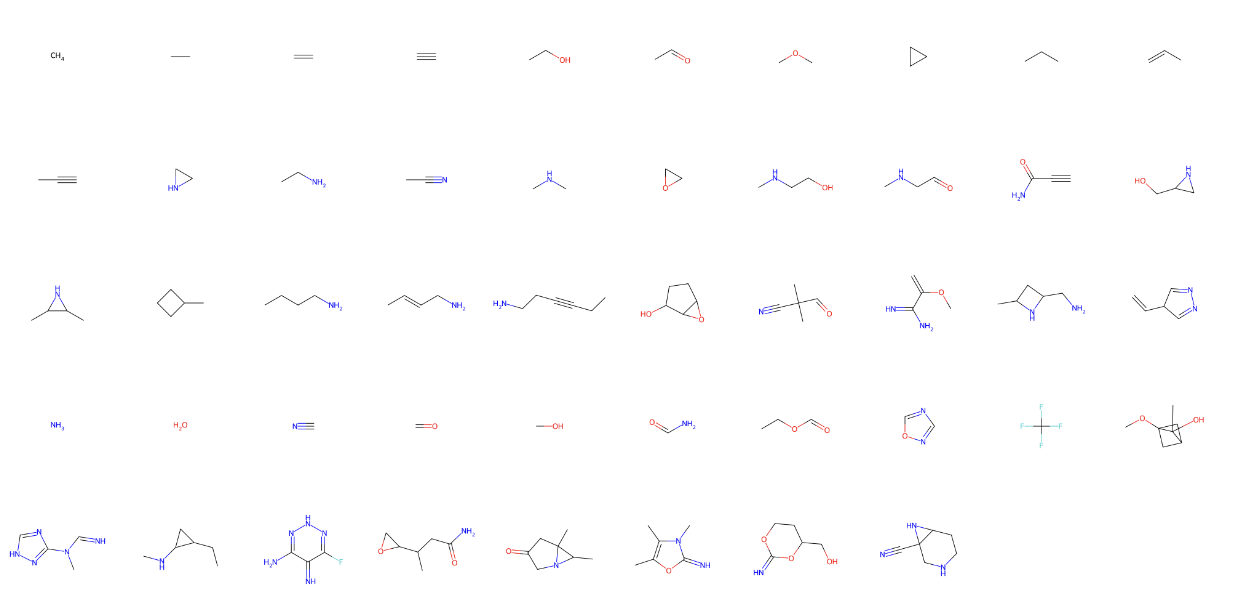}
\caption{Training molecules in the best model in terms of performance within our QM7 and QM9 dataset.}
\label{fig:si_rep7}
\end{figure}

\begin{figure}[hbt]
\centering
\includegraphics[width=0.95\textwidth]{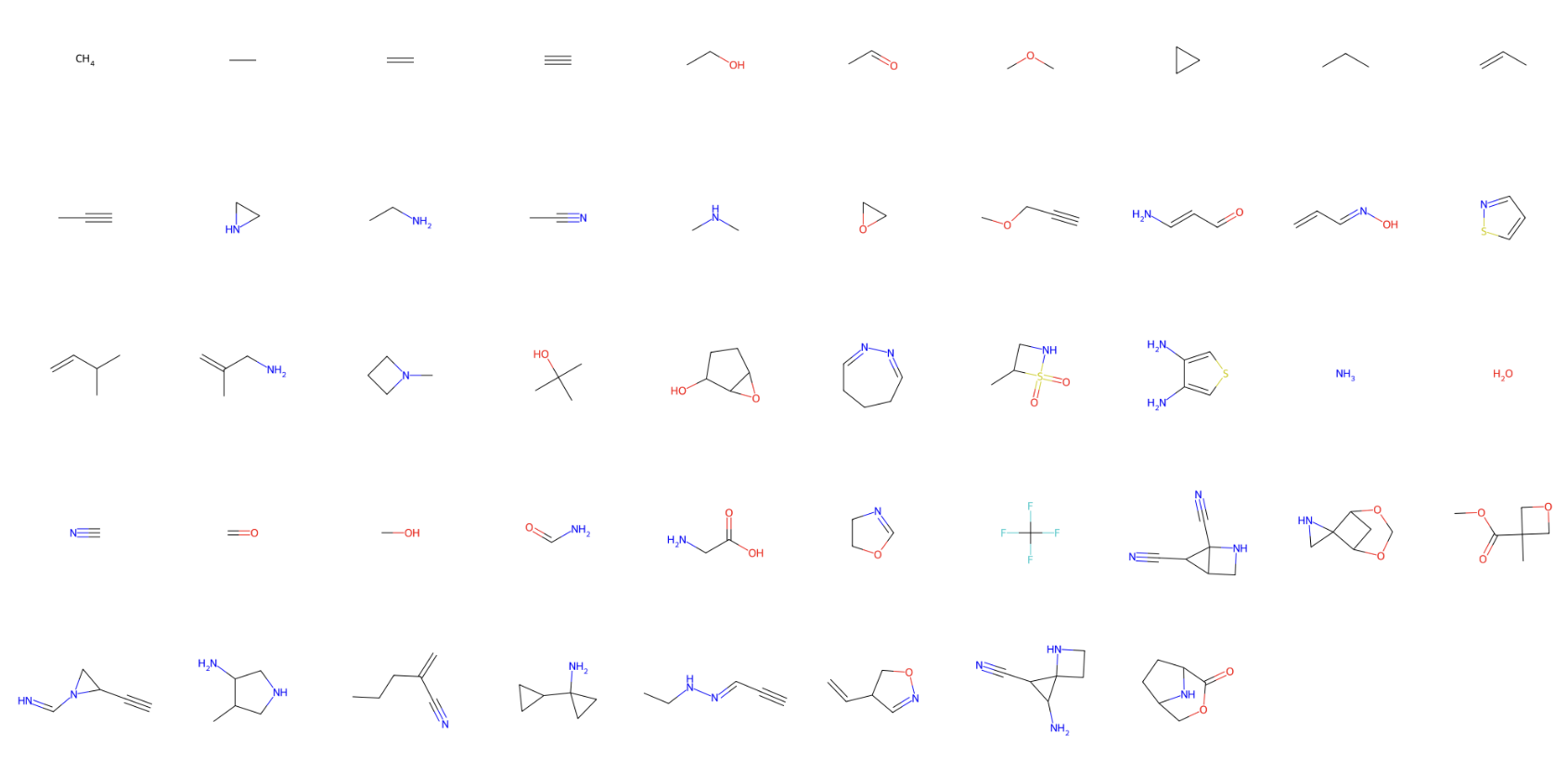}
\caption{Training molecules in the model used for predicting quinine and P3HT, chosen based on chemical intuition.}
\label{fig:si_rep4}
\end{figure}

\section{Frequency Points for Generating Dynamical Features}

\begin{figure}[hbt!]
\centering
\includegraphics[width=0.6\textwidth]{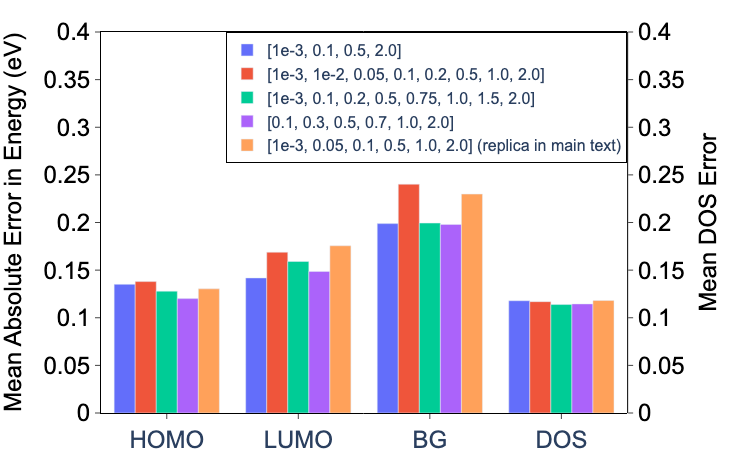}
\caption{Mean absolute errors of HOMO energy, LUMO energy, band gap (BG), and DOS on QM7 and QM9 datasets, by using 15 large molecules (each with 5 heavy atoms) as training data. Different sets of frequency points are used for generating dynamical features.}
\label{fig:dynfreq}
\end{figure}

We choose four other combinations of frequency points on which dynamical features are generated: $\omega=[10^{-3}, 0.1, 0.5, 2.0]$, $\omega=[0.1, 0.3, 0.5, 0.7, 1.0, 2.0]$, $\omega=[10^{-3}, 0.01, 0.05, 0.1, 0.2, 0.5, 1.0, 2.0]$, $\omega=[10^{-3}, 0.1, 0.2, 0.5, 0.75, 1.0, 1.5, 2.0]$, to compare against $\omega=[10^{-3}, 0.1, 0.2, 0.5, 1.0, 2.0]$ (the choice in main text, all $\omega$ values in a.u.). A benchmark on QM7 and QM9 datasets is shown in Fig.~\ref{fig:dynfreq}, where one replica is considered with 15 molecules (each with 5 heavy atoms) as training data. We find that the ML predictions on HOMO and DOS are not affected by different choices of frequency points, while the LUMO and band gap (BG) errors may be reduced by 0.02$\sim$0.03 eV by switching to a different set of frequency points. This test indicates the results reported in this paper are not sensitive to the choice of dynamical feature frequency points, although improvements can be achieved by more careful optimization in the future.

\end{document}